\documentclass[a4paper,fleqn]{cas-dc}

\usepackage[sort&compress,numbers]{natbib}

\newcommand{\wymiarsmall}{0.48\textwidth} %0.55 %Fig. 1
\begin{document}
\let\WriteBookmarks\relax
\def\floatpagepagefraction{1}
\def\textpagefraction{.001}
\shorttitle{Continuous unitary transformation approach to the Kondo-Majorana interplay}
\shortauthors{J. Bara\'nski, M. Bara\'nska, T. Zienkiewicz, J.Tomaszewska, K. J. Kapcia}
\title [mode = title]{Continuous unitary transformation approach to the Kondo-Majorana interplay}
\author[1]{Jan Bara\'nski}[%
                        orcid=0000-0002-0963-497X]
\ead{j.baranski@law.mil.pl}
\credit{Conceptualization, Methodology, Software, Formal analysis, Investigation, Resources, Writing - Original draft preparation, Writing - Review \& Editing, Visualization, Supervision, Project administration}

\author[1]{Magdalena Bara\'nska}[%
                        orcid=0000-0003-1964-6373]
\credit{Conceptualization, Methodology, Software,  Investigation, Resources, Data curation, Writing - Original draft preparation, Writing - Review \& Editing}

\author[1]{Tomasz Zienkiewicz}[%
                        orcid=0000-0002-7111-2590]
\credit{Validation, Formal analysis, Writing - Original draft preparation}

\author[1]{Justyna Tomaszewska}[%
                        orcid=0000-0001-6883-7235]
\credit{Validation, Resources}

\author[2,3]{Konrad~Jerzy Kapcia}[%
                    orcid=0000-0001-8842-1886]
\ead{konrad.kapcia@amu.edu.pl}
\credit{Software, Validation, Formal analysis, Investigation, Resources, Writing - Original draft preparation, Writing - Review \& Editing, Visualization, Supervision, Project administration, Funding acquisition}
\address[1]{Department of General Education, Polish Air Force University, ul. Dywizjonu 303 nr 35, 08521 Deblin, Poland} %problem with \k{e}
\address[2]{Institute of Spintronics and Quantum Information, Faculty of Physics, Adam Mickiewicz University in Pozna\'n, ul. Uniwersytetu Pozna\'{n}skiego 2, 61614 Pozna\'{n}, Poland}
\address[3]{Center for Free-Electron Laser Science CFEL, Deutsches Elektronen-Synchrotron DESY, Notkestr. 85, 22607 Hamburg, Germany}

\begin{abstract}
%\JB{
We analyze a setup composed of a correlated quantum dot (QD) coupled to one metallic lead and one end of topological chain hosting a Majorana zero mode (MZM). 
In such a hybrid structure, a leakage of the MZM into the region of the QD competes with the Kondo resonance appearing as a consequence of the spin-exchange interactions between the dot and the lead. 
In the work, we use the nontrivial technique called the continuous unitary transformation (CUT) to analyze this competition.
Using the CUT technique, we inspect the influence of the coupling between the QD and the chain on effective exchange interactions and calculate the resultant Kondo temperature.
\begin{itemize}
\item[$\ $]{\large{\textsc{Highligts:}}}
\item Quantum dot-based hybrid system is studied.
\item Majorana and Kondo physics interplay is investigated.
\item Continuous unitary transformation method is used.
\item Effective exchange interactions are determined.
\item Renormalization of model parameters is conducted.
\end{itemize}
\end{abstract}
%
%
%\begin{highlights}
%\item Quantum dot-based hybrid system is studied.
%\item Majorana and Kondo physics interplay is investigated.
%\item Continuous unitary transformation method is used.
%\item Effective exchange interactions are determined.
%\item Renormalization of model parameters is conducted.
%\end{highlights}
%
%
%
%\date{\today}
%
%
\maketitle
%
%
%\flushbottom
%
%

\section*{Introduction}
Hybrids comprising quantum dots (QD's) coupled to one end of topological chains hosting Majorana zero modes (MZMs) prove to be useful tools both for detection and for performing basic manipulations over 
these states \cite{Prada2017,Deng2016}. 
This is caused mainly by the tendency of MZMs to appear on the edges of 1D systems. 
Consequently, if a QD is tunnel-coupled to one edge of such wire, the zero energy state is induced in the spectrum of the QD. 
This effect is often referred to as ``leakage'' of the Majorana mode into the QD region \cite{Leakage2016,Kobiaka2019,Vernek,Baraski2016,Grski2018,MajekPRB2022,Baraski2021}. 
If a QD is additionally brought in contact with a metallic electrode another nontrivial feature appears exactly at the same energy level. 
Due to spin exchange between dot and lead, the virtual Kondo state pinned to the Fermi level is induced \cite{Kondo1964}.
Competition between these two processes is a many-body problem, which in general is not solvable exactly. 
One of the most reliable family of methods used to solve such problems are renormalization techniques. 
Using the numerical renormalization group theory (NRG), the Majorana-Kondo competition was analyzed, e.g., in single and double quantum dots \cite{Wojcik2014,MajekJMMM2022}.

In present work, we present a theoretical tool based on the continuous unitary transformation (CUT) \cite{Wegner1994,Glazek1994}. 
The CUT method was introduced by F. Wegner \cite{Wegner1994} and independently by K.G. Wilson with S. Glazek \cite{Glazek1994} and popularized mostly by S. Kehrein \cite{Kehrein1994,Kehrein1996,Kehrein2006}. 
Despite its advantages like retaining all degrees of freedom, i.e., the full Hilbert space or not requiring large computational powers, the method is not as popular as, e.g., numerical renormalization group.
The algorithm belongs to the techniques of the renormalization group and it allows to study correlation effects outside the perturbation scheme. 
At the technical level, the method develops the unconventional scaling of the entire Hilbert space, gradually separating low- and high-energy modes. 
The continuous Hamiltonian transformation is achieved by a set of scaling flow equations. 
For the setup considered in this work, scaling equations effectively decouple the QD from metallic lead (by elimination of hybridization terms), in return inducing additional (i.e., effective) terms, which were not present in the original Hamiltonian. 
One of these terms is directly related to mentioned spin exchange, other terms represent, e.g., density-density correlations between dot and lead. 
Solving the set of ordinary differential equations, we calculate the values of these interactions and calculate the dependence of Kondo temperature on coupling strength between the QD and the MZM.

\section{Model}
The system analyzed in this work
can be represented by the following Hamiltonian
\begin{eqnarray}
    \hat{H}=\hat{H}_{QD}+\hat{H}_{MQD}+\hat{H}_N+\hat{H}_{int},
    \label{H}
\end{eqnarray}
where 
\begin{eqnarray}
\hat{H}_{QD}\!=\!\sum_{\sigma}\! \epsilon_{\sigma} \hat{d}_{ \sigma}^{\dagger}\hat{d}_{ \sigma}+U_d \hat{n}_{\uparrow}\hat{n}_{\downarrow}
\label{QD}
\end{eqnarray}
represents single-level ($\epsilon_{\sigma}$) quantum dot with $\hat{d}^{\ }_{\sigma}$ ($\hat{d}^{\dagger}_{\sigma}$) being annihilation (creation) operators of $\sigma=\uparrow, \downarrow$ spin electrons, $U_d$ is the Coulomb repulsion between $\sigma=\uparrow$ and $\sigma=\downarrow$ electrons located on the QD and $\hat{n}_{\sigma}= \hat{d}^{\dagger}_{\sigma} \hat{d}^{\ }_{\sigma}$.
The metallic electrode is treated as a Fermi sea, i.e., $\hat{H}_{N}=\sum_{\bf{k},\sigma} \xi_{\bf{ k}} \hat{c}^{\dagger}_{ \bf{k}\sigma}\hat{c}_{ \bf{k}\sigma }$, where $\hat{c}^{\ }_{ \bf{k}\sigma}$ ($\hat{c}^{\dagger}_{ \bf{k}\sigma}$) are annihilation (creation) operators of $\bf{k}$ wave vector (and $\sigma$ spin) electrons with energy $ \xi_{ \bf{k}}=\epsilon_{\bf{k}}-\mu $ measured with respect to chemical potential $\mu$.
Low energy physics of a topological chain hosting pair of self-hermitan Majorana edge states ($\gamma^{\dagger}_1=\gamma_1$ and $\gamma^{\dagger}_2=\gamma_2$) and their coupling to the QD placed in vicinity of one edge state can be represented by the effective Kitaev model
\begin{eqnarray}
\hat{H}_{MQD}= i\epsilon_{m}\hat{\gamma}_{1}\hat{\gamma}_{2}+\lambda \left(\hat{d}_{ \uparrow}\hat{\gamma}_{1}+ \hat{\gamma}_{1}\hat{d}^{\dagger}_{\uparrow}\right),
\label{MQD1}
\end{eqnarray}
where $\epsilon_m$ is the overlap between two Majorana states and $\lambda$ represents tunneling rate of $\sigma=\downarrow$ spin electron between the QD and $\gamma_1$ edge state. 
Last part of the above equation represents electron hopping of $\downarrow$ electron from the QD to one edge of the chain and vice versa. 
The last part of Hamiltonian (\ref{H}) represents the electron hopping between dot and metallic lead
\begin{eqnarray}
    \hat{H}_{int}=\sum_{\bf{k}, \sigma} \left( V_{\bf{k} \sigma}  \hat{d}^{\dagger}_{ \sigma} \hat{c}^{\ }_{\bf{k} \sigma}+ \textrm{h.c.} \right),
    \label{Hint}
\end{eqnarray}
which in terms of the continuous unitary transformation approach will be treated as the interaction part. 
Using the standard method, we recast the Majorana operators in terms of ordinary fermion operators representation as 
$\hat{\gamma}_{1}=\left(\hat{f}+\hat{f}^{\dagger}\right)/\sqrt{2}$, and $\gamma_{2}=-\mathbf{i}\left(\hat{f}-\hat{f}^{\dagger}\right)/\sqrt{2}$.
This transformation implies, that (\ref{MQD1}) can be rewritten as
\begin{eqnarray}
\hat{H}_{MQD}&=&\epsilon_{m}\hat{f}^{\dagger}\hat{f}+t_{m} \left(\hat{d}_{\downarrow}^{\dagger}-\hat{d}_{ \downarrow}\right) \left(\hat{f}+\hat{f}^{\dagger} \right), 
\label{effective_MQD}
\end{eqnarray} 
where $t_m=\lambda/\sqrt{2}$.
The true unpaired Majorana edge states appear in the case of vanishing overlap of the edge states (i.e., $\epsilon_m=0$). 
In this work, we focus solely on such conditions.

The parameters of Hamiltonian (\ref{H}) like $V_{\bf{k} \sigma }$ or $\epsilon_{\sigma}$ are usually considered as spin-independent (i.e., $\epsilon_{\uparrow}=\epsilon_{\downarrow}$, $V_{\bf{k} \uparrow}=V_{\bf{k} \downarrow}$) unless additional magnetic field in the system is considered. 
However, we preserve  the spin index in eq. (\ref{H}). 
Hamiltonian (\ref{H}) and all corresponding parameters can be considered as special case (namely $l=0$) of renormalized $l$-space Hamiltonian $\hat{H}(l)$. 
Although mentioned parameters are \emph{apriori} exactly the same as for $l=0$, in general case for arbitrary $l$ parameter, it is not true  ($\epsilon_{\uparrow}(l) \neq \epsilon_{\downarrow}(l)$, $V_{\bf{k} \uparrow}(l) \neq V_{\bf{k} \downarrow}$(l)). 
Note that single spin coupling of QD to Majorana mode creates a spin imbalance in the system. 
Consequently, evolution of parameters $\epsilon_{\uparrow}$ and $\epsilon_{\downarrow}$ (as well as $V_{\bf{k} \uparrow} $ and $ V_{\bf{k} \downarrow}$) in $l$-space change differently.

\section{Outline of the method}

The main idea of continuous unitary transformation is based on a continuous process that step by step converts the input Hamiltonian, preserving full Hilbert space, to a diagonal or block-diagonal form by means of (usually) coupled scaling equations. 
In the initial stages of the continuous transformation, the most off-diagonal terms of the Hamiltonian (i.e., high-energy terms) undergo a continuous transformation, and then the states near the diagonal are rescaled. 
%\MB{
Keeping full energy information is important, e.g., when we are interested in correlation functions or non-equilibrium models.
%(?) \K{For me it is OK}}. 
The transformation of the Hamiltonian and other operators is generated by an anti-Hermitian operator, whose choice depends on the complexity and subtlety of the discussed problem \cite{Kehrein2006}.

In this section, we will outline the continuous unitary transformation for any Hamiltonian with the following structure
\begin{equation}
\hat{H} = \hat{H}_{0} + \hat{H}_{I} ,
\label{H0_HI}
\end{equation}
where $\hat{H}_{0}$ is the diagonal part (e.g., it could be kinetic energy
particles), and $\hat{H}_{I}$ corresponds to the off-diagonal part (describing the interactions
or disorders) or, as in the discussed case, the hybridization. 
In the continuous transformation $\hat{H}(l)=\hat{U}(l) \hat{H} \hat{U}^{\dagger}(l)$, 
the $l$-dependence ({\em flow}) of the Hamiltonian follows the differential equation 
\begin{eqnarray}
\frac{d\hat{H}(l)}{dl} = [\hat{\eta}(l),\hat{H}(l)],
\label{general}
\end{eqnarray} 
with the generating operator $\hat{\eta}(l) \equiv 
\frac{d \hat{{U}}(l)}{dl} \hat{{U}}^{-1}(l)$ (see \cite{Wegner1994,Glazek1994}). 
The choice of operator $\hat{\eta}(l)$ should provide
\begin{eqnarray}
\lim_{l\rightarrow \infty} \hat{H}_{I}(l) = 0 .
\label{limit}
\end{eqnarray}
One of the choices, proposed by S.Kehrein \cite{Kehrein1994}, is described by the following formula
\begin{eqnarray}
\hat{\eta}(l) = \left[ \hat{H}(l), \hat{H}_{I}(l) \right]. 
\label{KehrEta}
\end{eqnarray}

\section{Flow equations}

In our case, the generating operator should be chosen in such a way, that after renormalization ($l\rightarrow \infty$) the hybridization part in effective Hamiltonian should vanish ($V_{k\sigma}(l\rightarrow \infty)=0$). 
We determine the generating operator according to Eq. (\ref{KehrEta}) using hybridization part as $\hat{H}_I$.
The generating operator calculated in this way obeys the necessary condition of antihermitean structure $\hat{\eta}(l)
=\hat{\eta}_{0}(l)-\hat{\eta}_{0}^{\dagger}(l)$, where
\begin{eqnarray}
\label{gen1}
\hat{\eta}_{0}(l) &=& -\sum_{{\bf k} \sigma} V_{\bf{k} \sigma} \left ( \epsilon_{\sigma}-\xi_{\bf{k} \sigma}+U\hat{n}_{\bar{\sigma}}
\right)\hat{c}_{{\bf k}\sigma}^{\dagger} \hat{d}_{\sigma}  \\ 
\nonumber
&+&
 \sum_{{\bf k p} \sigma } V_{\bf{k} \sigma} V_{\bf p \sigma} \hat{c}_{{\bf k}
\sigma}^{\dagger} \hat{c}_{{\bf p}\sigma} + \epsilon_{\downarrow}t_m \left( \hat{d}_{\downarrow}^{\dagger}\hat{f}_{ }^{\dagger}
  +\hat{d}_{\downarrow}^{\dagger} \hat{f}  \right) 
\end{eqnarray}
In order to obtain the flow equations representing evolution of the model parameters such as $V_{\bf{k} \sigma},\epsilon_{\sigma}$,$t_m$ or $U_d$ in auxiliary $l$ space, we need to calculate $[\hat{\eta},\hat{H}]$ and compare the coefficients of the appropriate combinations of operators in the Hamiltonian. 
Let us introduce symbols:
$\tilde{\epsilon}_{\bf{k} \sigma} \equiv  \epsilon_{\sigma}-\xi_{\bf{k} }$ and
$\eta_{\bf{k} \sigma} \equiv  \tilde{\epsilon}_{\bf{k}\sigma}^2+\delta_{\sigma \downarrow} 2t_m^2 +U \langle n_{\bar{\sigma}} \rangle (2\tilde{\epsilon}_{\bf{k}\sigma}+U)$, where $\delta_{\sigma \sigma'}$ is the Kronecker delta. 
Using these symbols, we obtain 
\begin{eqnarray}
\label{etaH}
   [\hat{\eta},\hat{H}] &=& t_m \sum_{\bf{k}}V_{\bf{k} \downarrow}^2  (\hat{d}_{\downarrow}^{\dagger}-\hat{d}_{\downarrow})(\hat{f}^{\dagger}+\hat{f})\\ \nonumber
    &+& \sum_{\bf{k} \sigma}V_{\bf{k} \sigma}^2 \tilde{\epsilon}_{\bf{k} \sigma} \hat{d}^{\dagger}_{\sigma}\hat{d}_{\sigma}+ 2 \sum_{\bf{k} \sigma}V_{\bf{k} \sigma}^2U_d \hat{n}_{\sigma} \hat{n}_{\bar{\sigma}} \\ \nonumber
    &+&\sum_{\bf{k p}}\left(2  V_{\bf{p} \sigma}^2 -\eta_{\bf{k} \sigma}  \right)V_{\bf{k} \sigma} \left(\hat{c}^{\dagger}_{\bf{k} \sigma} \hat{d}_{\sigma}+\hat{d}^{\dagger}_{\sigma} \hat{c}_{\bf{k} \sigma}\right) \\ \nonumber
    &-& 2\sum_{\bf{k} \bf{p} \sigma}U_d V_{\bf{k} \sigma}V_{\bf{p} \sigma} \hat{c}_{\bf{k} \sigma}^{\dagger}\hat{c}_{\bf{p} \sigma}\hat{n}_{\bar{\sigma}} \\ \nonumber
    &+& 2\sum_{\bf{k} \bf{p} \sigma}U_d V_{\bf{k} \sigma}V_{\bf{p} \bar{\sigma}} \hat{c}_{\bf{k} \sigma}^{\dagger}\hat{c}_{\bf{p} \bar{\sigma}}\hat{d}^{\dagger}_{\bar{\sigma}}\hat{d}_{\sigma} + \hat{O}_3(l),
\end{eqnarray}
where $\hat{O}_3(l)$ are higher order terms. 
We note that apart from the combinations of operators occurring in original Hamiltonian (\ref{H}),
we obtain additional terms, which represent the effective density-density 
($c^{\dagger}_{\bf{k} \sigma}c_{\bf{p} {\sigma}}d^{\dagger}_{\bar{\sigma}}d_{\bar{\sigma}}$) 
and exchange  
($c^{\dagger}_{\bf{k} \sigma}c_{\bf{p} \bar{\sigma}}d^{\dagger}_{\bar{\sigma}}d_{\sigma}$) interactions between the dot and lead electrons.  
Such terms are also present in the standard Kondo Hamiltonian represented in terms of the Anderson impurity model with antiferromagnetic interaction $J<0$ \cite{Schrieffer1966}.
Let us rewrite the effective Hamiltonian including such terms:
\begin{eqnarray}
\label{Heff}
\hat{H}_{eff}(l)&=&\hat{H}(l)+\frac{1}{2}\sum_{\bf{k} \bf{p} \sigma}J_{\bf{k} p\sigma}(l) \hat{n}_{\bar{\sigma}}\hat{c}^{\dagger}_{\bf{k} {\sigma}}\hat{c}_{\bf{p} {\sigma}}\\ 
\nonumber &-& \frac{1}{2}\sum_{\bf{k} \bf{p}\sigma}J^{ex}_{\bf{k} \bf{p}}(l)\hat{d}^{\dagger}_{\sigma}\hat{d}_{\bar{\sigma}}\hat{c}^{\dagger}_{\bf{k} \bar{\sigma}}\hat{c}_{\bf{p} \sigma}+\hat{O}_3(l),
\end{eqnarray}
By comparing (\ref{Heff}) with flow equations (\ref{etaH}), one obtains the following set of differential equations (representing the $l$ renormalization of parameters):
\begin{eqnarray}
\label{EOM2}
\frac{d\epsilon_{\sigma}}{dl}&=& 2\sum_{\bf{k}} V_{\bf{k} \sigma}^2(\epsilon_{\sigma}-\xi_{\bf{k} \sigma}), \\
\frac{d U_d}{dl}&=& 2 \sum_{\bf{k} \sigma}V_{\bf{k} \sigma}^2U_d, \quad
\frac{d t_{m}}{dl}= \sum_{ \bf{k}} V_{\bf{k} \downarrow}^2 t_m,  \\
\label{EOM1}
\frac{d V_{\bf{k} \sigma}}{dl}&=& \left(2 \sum_{\bf{p}} V_{\bf{p} \sigma}^2 -\eta_{\bf{k} \sigma}  \right)V_{\bf{k} \sigma}.
\end{eqnarray}
Note that single spin coupling between the dot and the chain edge state creates a spin imbalance in the QD spectrum. Therefore, the interaction terms between two possible antiferromagnetic spin alignments (as well as the exchange interaction term) are  different in a general case. 
Therefore, in the effective Hamiltonian (\ref{Heff}), we assume arbitrary (non-identical) values of terms $J_{{\bf{k p}}\uparrow}, J_{{\bf{k p}} \downarrow}$ and $J_{{\bf{k p}}}^{ex}$.
The evolution of these additional terms is given by
\begin{eqnarray}
\frac{d J_{\bf{k p} \sigma}}{dl} =  -4 U V_{\bf{k} \sigma} V_{\bf{p} \sigma}, \quad
\frac{d J_{\bf{k p}}^{ex}}{dl} =  -4 U V_{\bf{k} \sigma} V_{\bf{p} \bar{\sigma}}
\label{EOMJ}
\end{eqnarray}
Using the Runge-Kutta method with initial conditions being non-renormalized parameters (i.e., $\epsilon(0)$, $U_d(0)$, etc.), we conduct the calculations until amplitude of hybridization parts $V_{\bf{k} \sigma}(l)$ will be reduced to negligible values. 
All effective terms ($J_{\bf{k} \bf{p} \sigma}, J_{\bf{k} \bf{p}}^{ex}$), which emerge during the flow calculations, have initial value zero as they are not present in the initial Hamiltonian (\ref{H}).

In the following part of the work, the dispersion relation is taken as a linear approximation: $\xi_{\bf{k}}=1-2|k|$ for  $k\in [-1,1]$  \cite{Kehrein1994}.  
Bandwidth $D$ of $\epsilon_{\bf{k}}$ is used as an energy unit.

\begin{figure}
        \centering
        \includegraphics[width=\wymiarsmall]{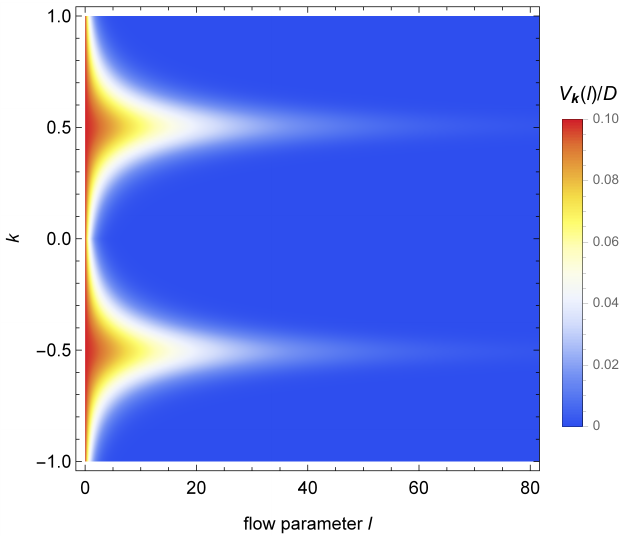}\\
	\includegraphics[width=\wymiarsmall]{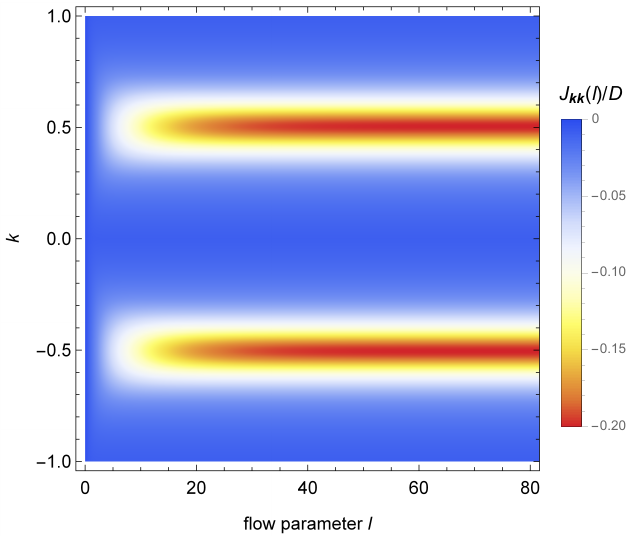}
	\caption{
            (top) Renormalization of the hybridization. 
            Decresing hybridization term $V_k$ (with $l$) obtained for $t_m(0)=0$, $\epsilon_{\sigma}(0) = -0.2D$, $V_{k}(0)=0.1D$, and $U_d (0)=0.4D$. 
            In the presented case, all hybridization terms are spin independent: 
            $ V_{k}  \equiv V_{k \uparrow} = V_{k \downarrow}$.
            (bottom) Renormalization of antiferromagnetic exchange $J_{ kk} \equiv J_{ kk\uparrow } = J_{ kk\downarrow }$ interactions 
            for the same model parameters. 
        }
	\label{figure1new}
\end{figure}

\section{Renormalization for vansihing dot-Majorana coupling}

For clarity let us first comment the renormalization of hybridization and exchange interactions in a case of vanishing dot-Majorana copuling term, i.e., $t_m(0)=0$. 
In Figure \ref{figure1new} we see that, upon decreasing hybridization $V_{k}(l)$ between the dot and the lead  (with increasing $l$), the magnitude of the effective exchange interaction $|J_{kk}(l)|$ increases. 
In the limit of $l \rightarrow \infty$, the hybridization vanishes completely and, in the effective Hamiltonian the QD is decoupled from the metallic reservoir. 
In contrary, the effective antiferromagnetic Kondo interactions ($J<0$) are induced. 
Note that, in a case of vanishing $t_m$, the renormalization of the hybridization occurs exactly the same for both spin components $V_{ {k} \uparrow}(l)=V_{k \downarrow}(l)$. 
This leads to induction of uniform  $J$ terms (for all three considered here interactions): $J_{{k p}\uparrow} = J_{{k p} \downarrow}=J_{{k p}}^{ex} \equiv J_{{k p}}$. 
This reproduces the results of the standard Kondo model for correlated impurity in vicinity of the Fermi sea \cite{Kondo1964}.

Following the single-step transformation performed by Schrieffer \cite{Schrieffer1966} for ${k}= {p} \simeq {{k}}_f$ (where $|{{k}}_{f}| = 1/2$ is the Fermi level in our case), the fully renormalized interaction coefficient should reach  
$ J = 2 |V_{{ {k}}_f}|^2 U_d / \left[ \epsilon_d \left( \epsilon_d+U_d \right) \right] $. 
In considered here half-filling conditions, i.e., $\epsilon_d=-U_d/2$, this expression reduces to %$J=-8\frac{|V_{kf}|^2}{U_d}$. 
$J=-8 |V_{{ {k}}_f}|^2 / U_d $. 
The resultant $J$ constant obtained within the CUT method $J(l \rightarrow \infty) \simeq-0.2D$,  (for the parameters from  Figure  \ref{figure1new}), is in full agreement with calculations performed by Schrieffer. 

\begin{figure}
        \centering        
        \includegraphics[width=\wymiarsmall]{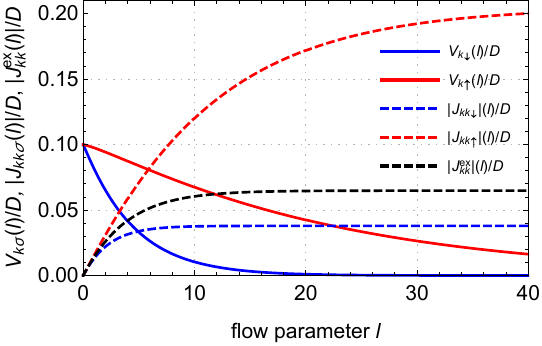}
	\caption{Flow of hybridization into exchange interactions.
 Solid red and blue lines, respectively, denote renormalization of hybridization terms $V_{k \uparrow}$, $V_{k \downarrow}$ at Fermi level ($k= k_f = 0.5$) obtained for initial dot-Majorana coupling $t_m(0)=0.3D$, $V_{k}(0)=0.1D$, and $U_d (0)=0.4D$. 
 Modules of exchange interactions $J_{kk \uparrow}$, $J_{kk \downarrow}$, and $J_{kk}^{ex}$ are represented by blue, red, and black dashed lines, respectively. 
        }
        \label{JVF}
\end{figure}

\begin{figure}
        \centering
        \includegraphics[width=\wymiarsmall]{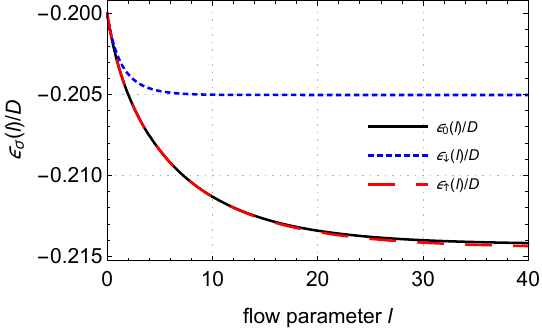}
	\caption{Renormalization of the dot energy.
            Energies $\epsilon_{\sigma}$ as a function of renormalization coefficient $l$. 
            Blue dotted and red dashed lines represent $\epsilon_{\downarrow}(l)$ and $\epsilon_{\uparrow}(l)$ obtained for initial dot-Majorana coupling $t_m(0)=0.3D$, $V_{k}(0)=0.1D$, and $U_d (0)=0.4D$. 
            For a comparison, the renormalization of $\epsilon_{0} \equiv \epsilon_{\downarrow} = \epsilon_{\uparrow}$ for the case of $t_m(0)=0$ (black solid line) is shown.}
	\label{epsRen}
\end{figure}

\begin{figure}
        \centering
        \includegraphics[width=\wymiarsmall]{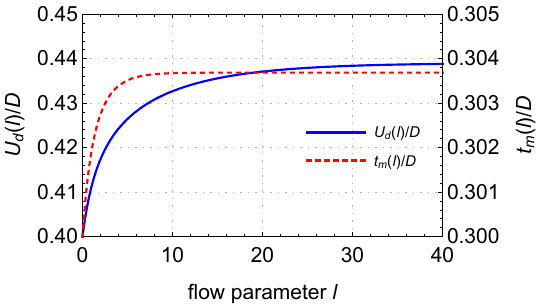}
	\caption{Renormalization of Coulomb interactions $U_d(l)$ (scale on the left) and  the dot-MZM coupling $t_m(l)$ (scale on the right).
        Blue solid line and red dashed line represent evolution of $U_d(l)$ and $t_m(l)$, respectively, obtained for initial value of dot-MZM coupling $t_m(0)=0.3$, $V_{k}(0)=0.1D$, and $U_d (0)=0.4D$. }
	\label{UtmRen}
\end{figure}

\section{Exchange and density-density correlations}
In this section all presented results are obtained for $t_m(0)=0.3D$, $\epsilon_{\sigma}(0) = -0.2D$, $V_{k}(0)=0.1D$ and $U_{d}(0)=0.4D$ if not indicated otherwise.
In Figure \ref{JVF}, we present the renormalization of hybridization and absolute values of $J$-terms obtained. 
Both hybridization terms  $V_{k \uparrow}$ and $V_{k \downarrow}$ are reduced by the magnitude of orders when $l$ parameter reach $80$.
Upon diminishing with $l$, hybridizations effective $J$-interactions are induced. 
We note that hybridization of $\sigma=\downarrow$ electrons (i.e. those who are tunnel coupled to MZM) is renormalized much faster compared to its opposite spin counterpart. Consequently, renormalization of $J$'s proceeds in a different way (and gives different final values). 
Term $J_{{k p} \uparrow}$, which is susceptible for changes of $V_{k \uparrow}$ (see Eq.~(\ref{EOMJ})) renormalizes longer and achieves a value comparable (slightly higher) to interaction amplitude obtained in the absence of the Majorana mode. 
In contrast, renormalization of $J_{{k p} \downarrow}$, is dependent on changes of $V_{\downarrow}$.
Thus, it renormalizes much faster, but it reaches only a fraction of $J_{{k p} \downarrow}$. 
Exchange interaction, which is dependent on both $V_{\uparrow}$ and $V_{\downarrow}$ reaches an intermediate value. 
These behaviors are also apparent in renormalization of dot energy levels $\epsilon_{\uparrow}$ and $\epsilon_{\downarrow}$ (see Figure \ref{epsRen}).
The renormalization of coupling strength between dot and chain $t_m(l)$ is neglible ($t_m(l)$ changes by about $1\%$ of initial value), but renormalization of $U_d$ interaction is about $10$~\%  (see Figure \ref{UtmRen}).

\begin{figure}
        \centering
    \includegraphics[width=\wymiarsmall]{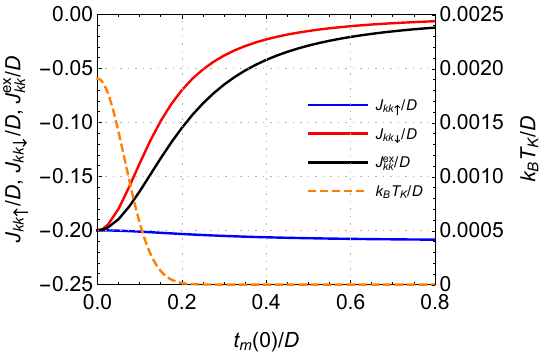}
	\caption{%
            Fully renormalized (for $l\rightarrow\infty$) exchange $J_{kk}^{ex}$  and density $J_{kk\sigma}$ interactions  obtained at the Fermi level ($|k|=1/2$) as a function of the dot-MZM coupling $t_m(0)$  (scale on the left). 
            The evolution of the Kondo temperature $T_K$ is indicated by the orange dashed line (scale on the right).}
	\label{3J}
\end{figure}

In Figure~\ref{3J}, we plot renormalized interaction $J$ terms for $k=p=k_f$ ($|k_f|=1/2$) versus dot-chain coupling $t_m$. 
We note that upon connecting the QD to the MZM, density-density interactions ($J_{{k p} \sigma}(l)$) for two possible antiferromagnetic alignments behave in the opposite way. 
Namely, the amplitude of the term $J_{{k p} \uparrow}$ representing alignment with $\downarrow$ electron on the dot and $\uparrow$ on the metallic side of the interface (cf. blue line in Figure \ref{3J}) is slightly strengthened. 
For the opposite spin alignment, amplitude of interactions $J_{{k p} \downarrow}$ is strongly reduced (cf. red line in Fig. \ref{3J}). 

Such inclination of the system for a given spin alignment can be understood by the effective Zeeman effect caused by coupling of the dot and the MZM mode. 
A similar effect has been reported by previous studies performed using the single-step Schrieffer-Wolff transformation \cite{Lutchyn}. 
It is also consistent with resultant spin degeneracy removal of the dot energy levels (cf. the renormalization of $\epsilon_{\uparrow}(l)$ and $\epsilon_{\downarrow}(l)$ in Figure~\ref{epsRen}).
We note that, in both cases, the obtained interactions obey $J_{ {k p }\sigma} < 0$, which still support antiferromagnetic spin alignment.

The crucial interaction leading to Kondo resonance is effective spin exchange between the dot and the lead ($J^{ex}_{ {k p}}$). 
Analyzing Figure~\ref{3J}, we note that coupling of the QD to the MZM has a detrimental effect on the exchange interactions. 
This indicates that leakage of the MZM into the QD region can effectively alter the Kondo resonance. 
This result is an in agreement with previous studies conducted in \cite{Lopez}. 
The Kondo temperature can be calculated using the Bethe Ansatz \cite{Tsvelick1983} as $k_B T_K = \frac{2}{\pi} D \exp\lbrace -\phi[2 \rho[\epsilon_{{k_f}}J^{ex}_{{k_f} k_f}(l=\infty) ] \rbrace$, where $\rho(\epsilon_{{k_f}})$ is the density of states at the Fermi level and $\phi(y) \simeq |y|^{-1}-0.5 \ln |y|$. 
Fig. \ref{3J} presents that the Kondo temperature decreases gradually upon connecting QD and MZM. 

\section{Conclusions}
We performed a continous unitary transformation for the system composed of a correlated quantum dot coupled to one metallic electrode and one end of a nanowire hosting a Majorana quasiparticle. 
We showed that, due to connection of the QD to the MZM, the effective Zeeman field removes the spin degeneracy of the QD energy levels. 
Consequently, the interactions between dot and lead electrons renormalizes in a different manner, and preferable spin alignment appears in the system. 
Competition of MZM leakage and spin-exchange interaction between the dot and the lead reduces the effective amplitude of exchange interactions leading to a significant lowering of the Kondo temperature. 
Finally, we showed that presented procedure, in the case of vanishing dot-chain coupling, reproduces the well-known result obtained using a single-step Schrieffer-Wolff technique \cite{Schrieffer1966}.

\section*{Acknowledgements}

We thank Tadeusz Doma\'nski for very fruitful discussions and careful reading of the manuscript. J.T. thanks the 
Ministry of National Defence of Poland for funding in the frame of the SZAFIR project (DOB-SZAFIR/01/B/023/01/2020).
K.J.K. thanks the Polish National Agency for Academic Exchange for funding in the frame of the Bekker programme (PPN/BEK/2020/1/00184). 

\section*{Declaration of Competing Interest}
The authors declare that they have no known competing financial interests or personal relationships that could have appeared to influence the work reported in this paper.
The funders had no role in the design of the study; in the collection, analyses, or interpretation of data; in the writing of the manuscript, or in the decision to publish the results.

\printcredits

\bibliographystyle{elsarticle-num}

\bibliography{biblioPM21}
\end{document}